\documentclass[useAMS,usenatbib]{mn2e}

\usepackage{times,graphicx,amssymb,natbib}

\title[Native Synthetic Imaging of SPH density fields using MCRT]{Native Synthetic Imaging of Smoothed Particle Hydrodynamics density fields using gridless Monte Carlo Radiative Transfer}
\author[Duncan Forgan and Ken Rice]{Duncan Forgan $^{1}$\thanks{E-mail:
dhf@roe.ac.uk} and Ken Rice$^{1}$\\
$^{1}$Scottish Universities Physics Alliance (SUPA), Institute for Astronomy, University of Edinburgh, Blackford Hill, Edinburgh, EH9 3HJ, Scotland, UK \\
}
\begin{document}

\date{Accepted 2010 April 13.  Received 2010 April 10; in original form 2010 January 29}

\pagerange{\pageref{firstpage}--\pageref{lastpage}} \pubyear{0000}

\maketitle

\label{firstpage}

\begin{abstract}

\noindent An algorithm for creating synthetic telescope images of Smoothed Particle Hydrodynamics (SPH) density fields is presented, which utilises the adaptive nature of the SPH formalism in full.  The imaging process uses Monte Carlo Radiative Transfer (MCRT) methods to model the scattering and absorption of photon packets in the density field, which then exit the system and are captured on a pixelated image plane, creating a 2D image (or a 3D datacube, if the photons are also binned by their wavelength).  The algorithm is implemented on the density field directly: no gridding of the field is required, allowing the density field to be described to an identical level of accuracy as the simulations that generated it.

Some applications of the method to star and planet formation simulations are presented to illustrate the advantages of this new technique, and suggestions as to how this framework could support a Radiative Equilibrium algorithm are also given as an indication for future development.

\end{abstract}


\section{Introduction}

\noindent The physics of dust is of paramount importance when modelling star and planet formation: dust signatures are seen in both molecular clouds and circumstellar discs, and are a crucial component in their observed properties over a wide range of wavelengths.  Indeed, the existence of dust is essential if planets are to be formed inside these discs.  The presence of dust has important effects on the radiative signatures that can be detected by astronomers: dust will scatter and polarise at short wavelengths, as well as reprocessing this radiation to longer wavelengths.  The nature of the scattering and polarisation will depend on the geometry of both the system and the dust in the system, as well as the physical properties of the dust itself (composition, morphology, grain size, etc).

In recent years, telescopes/networks such as HST, SCUBA, MERLIN and Spitzer have efficiently probed stellar systems from the IR to the radio, with future ground and space-based missions such as Herschel, ALMA, SCUBA II, JWST, SPICA and e-MERLIN improving the quality of this data.  Astronomers will soon be faced with a wealth of new, high-fidelity astronomical data on stellar objects across a wide spectral range, allowing detailed studies of the evolution of dusty disc systems, in particular the interplay between disc dust and disc gas.  For numerical simulations to inform these observations, the simulation of radiative transfer (RT) in these systems must be pursued, so that imaging of numerical simulations can provide theoretical insights to observations. 

Monte Carlo Radiative Transfer (MCRT) has become a popular tool in simulating and imaging astrophysical systems \citep{Whitney_and_Hartmann_92,Wood_et_al_96,MCFOST}.  The technique involves the tracking of \emph{photon packets} in the medium, allowing them to scatter and absorb in the medium (as well as attaining a non-zero polarisation).  The photons are tracked until they escape the medium, and can then be captured on an image plane.  Traditionally, the method requires the density field to be defined to as high an accuracy as possible - this is usually achieved by gridding the density field in 3 dimensions.

Smoothed Particle Hydrodynamics (SPH) is a Lagrangian method which represents a fluid by a particle distribution \citep{Lucy,Gingold_Monaghan}.  Each particle is assigned a mass, position, internal energy and velocity.  From this, state variables such as density can then be calculated at any position in the system by interpolation - see reviews by \citet{Monaghan_92, Monaghan_05}. It has been successful in modelling astrophysical systems of various scales and geometries, from protostellar systems to galactic systems and beyond.  Its key advantage is the ability to follow the change of density through many orders of magnitude adaptively.  Gravity can be included in SPH calculations, and optimised using traditional hierarchical tree methods \citep{Hernquist_and_Katz_89}. The formalism can also be modified to simulate magneto-hydrodynamics (MHD) \citep{Hosking_Whitworth_04,Price_Monaghan_SPMHD2,Price_Monaghan_SPMHD3,Price_and_Bate_07}.

Radiative physics is crucial to the simulation of any astrophysical fluid.  Radiative transfer in SPH has a long history, with efforts ranging from simple parametrisations for the radiative cooling time (e.g. \citealt{Ken_1}), optical-depth dependent radiative cooling (e.g. \citealt{Stam_2007}), through to flux-limited diffusion models (e.g. \citealt{WB_1,Bastien_diffusion,Viau_et_al_06,WB_2, Mayer_et_al_07, intro_hybrid}).  While these algorithms are extremely well-suited to calculating gas temperatures during runtime, they lack the precision and insight offered by radiative transfer techniques which do not average over frequency, such as MCRT.

MCRT techniques can be applied to the SPH density field, and have been on several occasions.  Early attempts began by binning the particle distribution onto a grid - however, the choice of geometry strongly influences the final gridded field, and often adaptive meshes are required to correctly represent the matter distribution \citep{Oxley_Woolfson_2003,Kurosawa_et_al_04,Stam_MCRT,Bisbas_et_al_09}.  Later efforts have utilised ray-tracing directly in SPH fields \citep{Kessel_Deynet_and_Burkert_00,SPHRAY,TRAPHIC}, allowing the full power of the SPH formalism to be applied to the calculation of optical depths. However, these methods currently only calculate optical depths along the ray for the purpose of photoionisation, etc, and do not account for detailed scattering and polarisation, which are important in imaging small-scale systems such as protoplanetary discs. 

This paper introduces an algorithm for imaging SPH simulations of star and planet formation directly using MCRT, without requiring gridding, and including scattering and polarisation.  It utilises the same techniques that a traditional gridded MCRT code uses, with the advantage that it can trace rays in the density field with the same adaptive capability as SPH, such that it can model radiative effects with at least the same resolving power.  The paper is organised as follows: section \ref{sec:method} will outline the algorithm, sections \ref{sec:applications} and 4 will describe some applications of the technique to imaging numerical simulations, and section \ref{sec:conclusions} will summarise the results of this work.

\section{Method} \label{sec:method}

\subsection{Monte Carlo Radiative Transfer}

\noindent For completeness, a summary of MCRT is given here.  \emph{Photon packets} are emitted from sources (these can be point sources or diffuse emission from the density field itself).  These packets (hereafter referred to simply as photons) then travel through the medium, and interact with the medium stochastically, reproducing the statistical scattering properties of the medium.  The exact formalism is summarised below.

\subsubsection{Emission of Photons}

\noindent Photons can be emitted either from a central source (e.g. a protostar), or as diffuse emission from the density field itself, which is only done if the temperature structure of the diffuse component is known (e.g. as output from a simulation). \emph{Radiative equilibrium} methods in MCRT (e.g. \citealt{Bjorkman_Wood_MCRE}) can calculate the temperature structure directly from any luminosity source (e.g. stellar, accretion, external radiation fields).  However, as the SPH simulations used in this work produce temperatures for each SPH particle self-consistently, the system is assumed to already be in temperature equilibrium.  It is also assumed that the dust is thermally coupled to the gas, and that $T_{dust} = T_{radiation}$.  This is suitable for most purposes, except where significant stellar irradiation  dominates the radiation field (not modelled in the simulations described in this paper).  However, as has been already said, radiative equilibrium techniques can be used to sidestep this issue: an implementation of radiative equilibrium in SPH fields is discussed in a later section.

It is reasonable to (initially) assume all objects emit according to a blackbody spectrum, which then gives:

\begin{equation} L_{star} = 4 \pi^2 R^2_{s} \int_{\nu_{min}}^{\nu_{max}} B_{\nu}(T_{s}) d\nu \label{eq:L_point}\end{equation}
\begin{equation} L_{gas} = M_{gas} \int \left[\int_{\nu_{min}}^{\nu_{max}} \epsilon_{\nu} (\mathbf{r}) d\nu \right] d\mathbf{r}, \label{eq:L_diffuse} \end{equation}

\noindent for source and diffuse emission respectively (given the frequency range \([\nu_{min}, \nu_{max}]\) of interest), where $\epsilon_{\nu} (\mathbf{r})$ is the emissivity interpolated from the SPH particle field, i.e.

\begin{equation} \epsilon_{\nu} (\mathbf{r}) = 4 \pi \sum_{j} \frac{\kappa_{\nu} B_{\nu} (T_j)m_j}{\rho_j} \textbf{W}(\textbf{r} - \textbf{r}_j,h), \end{equation}

\noindent where the sum over $j$ indicates a sum over nearest neighbours, $\kappa_{\nu}$ is the absorptive opacity, $T_j$ is the temperature of each SPH particle, $m_j$ and $\rho_j$ are the masses and densities respectively, and $\mathbf{W}$ is the smoothing kernel (see section \ref{sec:tau_SPH} for more detail on the SPH formalism).

There is an extra subtlety regarding the frequency distribution of photons emitted from the gas.  The frequency distribution will depend on the local emissivity, which as shown above is an interpolative sum.  While in theory the full form of equation (\ref{eq:L_diffuse}) should be used to calculate gas luminosity (and the frequency distribution of the photons it emits), this paper approximates the sum using the contribution from one particle only, i.e.

\begin{equation} L_{gas} = \sum_{i} L_{i} = \sum_{i} 4 \pi M_{i} \int_{\nu_{min}}^{\nu_{max}} \kappa_{\nu} B_{\nu} (T_i) d\nu. \end{equation} 

\noindent This saves computational expense, and is sufficiently accurate for the examples in this paper,  bearing in mind a) that the approximation breaks down only in the inner regions of the disc, which are already under-resolved for other reasons (see section \ref{sec:res}) and b) the effective resolution of the images is too low for this effect to be significant.

The luminosity of each object (whether pointmass or SPH particle) defines how many photon packets are emitted from that object using

\begin{equation} N_{\gamma,object} = N_{\gamma,tot}\left(\frac{L_{object}}{L_{tot}}\right) \label{eq:nphotons}\end{equation} 

\subsubsection{Absorption/Scattering of Photons}

\noindent In order to correctly reproduce the scattering of photons in the medium, the cumulative distribution function (CDF) of interaction 

\begin{equation} F(\tau) = 1 - e^{-\tau} \end{equation}

\noindent must be reproduced.  In essence, this implies sampling an optical depth using

\begin{equation} \tau_{scatter} = -1-\log(1-\zeta) \end{equation}

\noindent where \(\zeta\) is a random number between 0 and 1.  The photon will travel a distance \(L\) along a constant vector path (hereafter \emph{ray}), such that

\begin{equation} \tau_{scatter} = \int^L_0 \rho\, \chi_{\nu} \, d\ell,  \label{eq:tau}\end{equation}  

\noindent (where $\chi_{\nu}$ is the total opacity), and then be either absorbed or scattered.  The majority of computation required in any MCRT code is the solution of equation (\ref{eq:tau}).  As the density field of an astrophysical system is in general non-trivial, analytic solutions cannot be applied, and numerical procedures must be used.  The algorithm must be able to calculate the optical depth along the ray for arbitrary distances - this requirement is what demands the most CPU time.  To simplify matters,  most MCRT codes use gridding as a means of defining the density field simply in space. 

Once the photon has reached the scattering location, its direction after scattering must be calculated.  The angle of scattering is defined by the scattering matrix M which acts on the Stokes vector \(S=(I,Q,U,V)\), where \(I\) is the intensity, \((Q,U)\) are the linear polarisations at 45$^{\circ}$ to each other, and V is the circular polarisation:

\begin{equation} S' = R(\pi - i_2)\, M\, R(-i_1) \, S. \label{eq:Stokes} \end{equation}

\noindent The \(R\) matrices are Mueller matrices, which describe rotations to and from the observer's frame.  They are defined as:

\begin{equation} R(\psi) = \left[ 
\begin{array}{c c c c}
1 & 0 & 0 & 0 \\
0 & \cos{2\psi} & \sin{2\psi} & 0 \\
0 & -\sin{2\psi} & \cos{2\psi} & 0 \\ 
0 & 0 & 0 & 1 \\
\end{array}  \right]
\end{equation} 

\noindent The scattering matrix M is dependent on the dominant source of scattering in the medium.  It can be expressed as a function of several scattering parameters, \(M_i\):

\begin{equation} M(\Theta) = a\left[
\begin{array}{c c c c}
M_1 & M_2 & 0 & 0 \\
M_2 & M_1 & 0 & 0 \\
0   & 0   & M_3 & -M_4 \\ 
0   & 0   & M_4 & M_3 \\
\end{array}  \right] 
\end{equation}

\noindent The scattering angle \(\Theta\) and azimuthal angle \(\phi\) are sampled randomly from the scattering matrix, using the cumulative distribution functions: 

\begin{equation} F(\Theta) = \frac{\int^{\Theta}_{0}M_{1}\sin\Theta' d\Theta'}{\int^{\pi}_{0}M_{1}\sin\Theta' d\Theta'} \label{eq:Theta}\end{equation}

\begin{equation} F_{\Theta} (\phi) = \frac{1}{2\pi}\left(\phi - \left(\frac{M_1-M_2}{M_1+M_2}\right)\frac{P}{2} \sin 2\phi \right) \end{equation}

\noindent Where 

\begin{equation} P = \frac{\sqrt{Q^2 + U^2}}{I} \end{equation}

\noindent In general, whether the photon is absorbed or scattered, the total number of photons is conserved by forcing absorbed photons to be immediately re-emitted.  As MCRT deals with \emph{photon packets}, energy can be conserved by reducing the energy of each photon packet after a scattering/absorption event by a factor equal to the local albedo.  This equates with the concept of a fraction of the photons in the packet being absorbed by the medium, and the complementary fraction being scattered.  Other methods can also be used, e.g. ``killing" a photon if the local albedo is less than a randomly sampled value, which saves computing emission from low-intensity packets.  However, this work uses the former method rather than the latter.  In the following work, the absorbed photons do not affect the local temperature structure - it is assumed that the equilibrium temperature is known (having been produced by the SPH simulation).

\subsubsection{Imaging}

\begin{figure}
\begin{center}
\includegraphics[scale = 0.25]{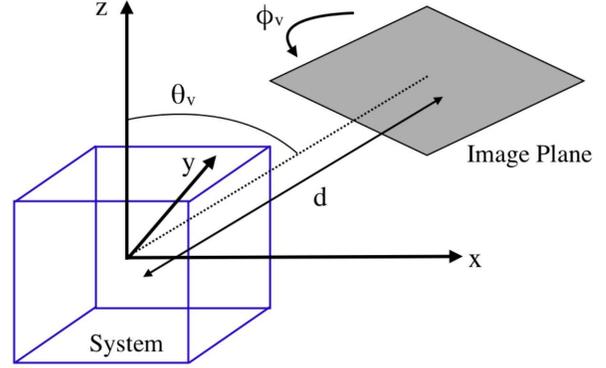}
\caption{Defining an image plane \label{fig:image}.}
\end{center}
\end{figure}

\noindent When photons exit the medium, they are captured on an image plane, oriented at user-specified angles \(\theta_v, \phi_v\) to the system, at a fixed distance $d$ (see Figure \ref{fig:image}).  They are then binned by their \((x,y)\) coordinates on this plane to provide a pixelated image, averaged over solid angle (analogous to imaging in a CCD).  If spectra are of interest to the user, then these can also be obtained by binning in \(\lambda\) (or indeed, an entire datacube can be obtained by binning in all three).   ``Classic" MCRT methods do not specify a single viewing angle, and instead bin the photons over several lines of sight, which allow the construction of a series of image planes from one simulation.  However, such ``multi-plane" simulations will have to run many more photons than a ``one-plane" simulation, to maintain a comparably low level of random error associated with each pixel\footnote{This is also a factor when comparing one-plane simulations with different pixel resolutions.}.  For this reason the code used in this work adopts the ``one-plane" imaging scheme.

This numerical apparatus requires the code to be run once for every desired orientation of the image plane.  It is possible to avoid this issue by defining a series of image planes around the system, but the numerical resolution of the Monte Carlo method is reduced as the pixel to photon ratio is increased, so this is in general avoided to prevent significant computational expense.  

The image plane size is defined in physical units, e.g. an image that captures features a distance of $r$ au from the centre of the system.  This prevents photons from features outside $r$ (or photons which escape with a vector which does not intersect the plane) from being recorded in the final image.  Having specified $r$ and the distance $d$, this sets the angular size of the image.  Coupled with information regarding the angular resolution of the telescope being simulated, this defines how well resolved the image is, and the calculated flux.  In lieu of more sophisticated telescope simulation, the image is simply smoothed pixel by pixel with a Gaussian with width equal to the angular resolution.  While obviously far short of a true synthetic image, it remains a useful first-order approximation for this work.  

\subsection{Implementing Gridless MCRT}

\subsubsection{Optical Depths in an SPH density field \label{sec:tau_SPH}}

\noindent As has been said previously, the key component of any MCRT code (and the source of the greatest computational burden) is the calculation of optical depths.  This calculation requires a specification of the density field at all locations, which can be given by the SPH formalism.  The definition of a scalar field \(A(\textbf{r})\) using an SPH particle distribution is

\begin{equation} A(\textbf{r}) = \sum_{j} \frac{A_jm_j}{\rho_j} \textbf{W}(\textbf{r} - \textbf{r}_j,h) \end{equation}

\noindent where \(A_j\) is the value of \(A\) at the location of particle \(j\), \(m_j\) is the mass of particle \(j\), \(\rho_j\) is the density of particle \(j\), and \(\textbf{W}\) is the smoothing kernel (or interpolating kernel).  The summation is typically over \(N_{neigh}\) nearest neighbour particles to the location \(\textbf{r}\).  The \emph{smoothing length} of particle \(j\), \(h_j\), is defined such that a sphere of radius \(2h_j\) will contain the \(N_{neigh}\) nearest neighbour particles to \(j\).  For example, to calculate density, substitute \(A=\rho\):

\begin{equation} \rho(\textbf{r}) = \sum_{j} m_j \textbf{W}(\textbf{r} - \textbf{r}_j,h). \label{eq:rhocalc} \end{equation}

\noindent The sphere that contains the \(N_{neigh}\) nearest neighbours (i.e. a sphere of radius \(2h_j\)) is referred to as the \emph{smoothing volume}.There is a subtlety to equation (\ref{eq:rhocalc}) that should be noted, relating to which value of \(h\) to use.  There are now two means by which to estimate density: the first is the so-called ``gather'' method, where the smoothing length \(h=h_i\) is defined for the location \(r_i\), and

\begin{equation} \rho(\textbf{r}_i) = \sum_j m_j \textbf{W}(\textbf{r}_i - \textbf{r}_j,h_i) \label{eq:gather} \end{equation}

\noindent Where the index \(j\) refers to all particles which are contained within a radius \(2h_i\) of the location \(r_i\). \\

\noindent The second method (which is used in this work and in {\small SPHRAY} \citealt{SPHRAY}) is the ``scatter'' method.  The smoothing length \(h=h_j\) is used - in this formalism, the density at any one location is calculated by adding the contributions from particles whose smoothing volume intersects the location:

\begin{equation} \rho(\textbf{r}_i) = \sum_{j} m_j \textbf{W}(\textbf{r}_i - \textbf{r}_j,h_j). \label{eq:scatter} \end{equation}

\begin{figure}
\begin{center}
\includegraphics[scale = 0.25]{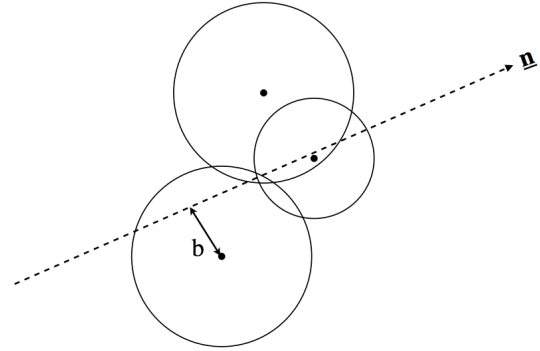}
\caption{Illustrating the ``scatter'' method.  Particles only contribute to the density along the ray if their smoothing volume intersects it\label{fig:scatter}.}
\end{center}
\end{figure}

\noindent In the context of ray tracing, the density along the ray is affected only by particles with smoothing volumes that intersect it (see Figure \ref{fig:scatter}).  By determining which particles intersect the ray, the rest of the particle distribution can be ignored for the purposes of calculating optical depth, reducing computational expense (whereas with the gather method, the ensemble of particles contributing to the calculation changes significantly with position, and requires the inclusion of a larger subset of SPH particles to perform the calculation). \\

\noindent Using the scatter method, the column density \(\Sigma\) along the ray is 

\begin{equation} \Sigma = \int^{L} _{0} \rho(\textbf{r}) d \ell = \int^{L} _{0}  \sum_{j=1}^N \left[m_j \textbf{W}(\textbf{r}_i - \textbf{r}_j,h_j)\right] d \ell \end{equation}

\noindent Which can be rearranged to give

\begin{equation} \Sigma =  \sum_{j=1}^N \left[  \int^{L} _{0}  m_j \textbf{W}(\textbf{r}_i - \textbf{r}_j,h_j) d \ell \right] \label{eq:sigma}\end{equation}

\noindent The integral is now decomposed into \(N\) integrals, where \(N\) is the number of particles intersected by the ray.  Each integral is defined by the impact parameter \(b\) (see Figure \ref{fig:scatter}).  The calculation itself can be performed for a smoothing volume of \(h=1\), and scaled upwards (this is due to the construction of the smoothing kernel).  The entire optical depth calculation has been decomposed into the repetition of a single algorithm for calculating the optical depth through a single smoothing volume.  This calculation will now be expounded. \\

\subsubsection{The Optical Depth Calculation for A Single Particle}

\begin{figure}
\begin{center}
\includegraphics[scale = 0.25]{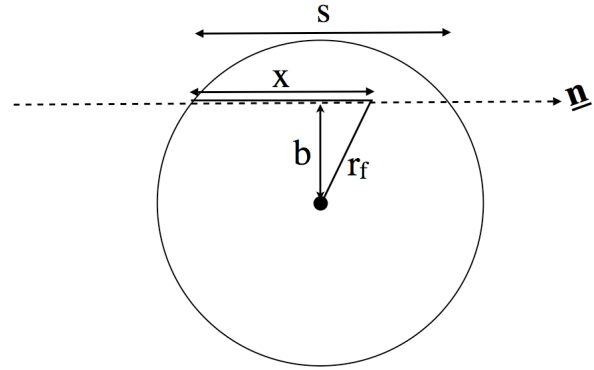}
\caption{Optical Depths through a single smoothing volume \label{fig:sphere}.}
\end{center}
\end{figure}	 

\noindent Consider Figure \ref{fig:sphere}.  The ray (with direction vector \(\textbf{n}\)) intersects the sphere with impact parameter \(b\).  If the ray penetrates a distance \(x\) into the sphere (out of a total possible distance \(s\)), then the integral can be defined analytically, given the functional form of \(\textbf{W}\).  defining 

\begin{equation}\tilde{r} = r/h  \end{equation}
\begin{equation}\tilde{b} = b/h \end{equation}
\begin{equation} \tilde{x} = x/h \end{equation}

\begin{equation} r_f = \sqrt{(s/2 -x)^2 +b^2} \end{equation}

\noindent (and \(\tilde{r}_f = r_f/h\)), the integral in equation (\ref{eq:sigma}) becomes:

\begin{equation} I = \left\{
\begin{array}{l l }
 \int^{2}_{\tilde{r}_f} W(\tilde{r}') d\tilde{r}' & \quad x< s/2 \\
 \int^{2}_{\tilde{b}} W(\tilde{r}') d\tilde{r}' +  \int^{\tilde{r}_f}_{\tilde{b}} W(\tilde{r}') d\tilde{r}' & \quad x> s/2 \\
\end{array} \right. \end{equation} 

\noindent The column density through a smoothing volume (for any impact parameter and any distance into the sphere) can now be calculated; multiplying by an opacity then provides an optical depth.  Typically, \(\textbf{W}\) is constructed using cubic splines \citep{Monaghan_92}.  The kernel used in this work is

\begin{equation} W(\tilde{r}) = \left\{
\begin{array}{l l }
 1 - \frac{3}{2} \tilde{r}^2 + \frac{3}{4} \tilde{r}^3 & \quad \tilde{r}< 1 \\
 \frac{1}{4} (2-\tilde{r})^3& \quad 1< \tilde{r}< 2  \\
 0 & \quad \tilde{r} > 2 \\
\end{array} \right. \end{equation} 

\noindent This provides compact support (i.e. it reduces to zero outside the smoothing volume), and is simple to integrate.  

\subsubsection{Ray Tracing in an SPH density field}

\noindent With a prescription for calculating optical depth for a single sphere in place, a scheme for calculating ray/sphere intersections must be constructed.  To this end, the code creates a data object called a \emph{raylist}, which stores (in order of intersection) all particles that the ray (given its origin and direction vector) will intersect.  Once the list is created, the optical depth can be calculated quickly using equation (\ref{eq:scatter}).

\begin{figure}
\begin{center}
\includegraphics[scale = 0.25]{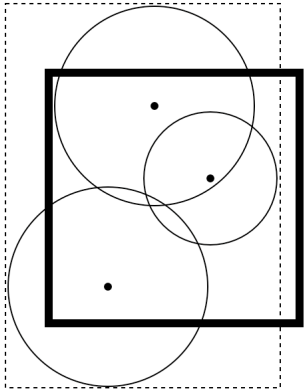}
\caption{Schematic of an Axis Aligned Bounding Box (AABB).  For a given set of occupants of a tree cell (solid line), the AABB (dashed line) is set to the minimum dimensions required to completely contain their smoothing volumes. \label{fig:aabb}}
\end{center}
\end{figure}

The construction of the raylist must be computationally efficient for the code to be effective.  The procedure is similar to that implemented by \citet{SPHRAY} in  {\small SPHRAY}; the code constructs an octree to spatially index the particles efficiently (as there may be density changes over several orders of magnitude).  The cells either contain child cells, or particles (the \emph{leaf cells}).  The tree is constrained to have a maximum number of particles in each leaf.  All cells have an associated Axis Aligned Bounding Box (AABB), which is the minimum box size, aligned to the three cartesian axes, to contain all the smoothing volumes of the particles in the cell (see Figure \ref{fig:aabb}).  These AABBs are necessary as tree nodes may contain a particle, but not its entire smoothing volume 

This allows the determination of intersections between the ray and the cells (or more correctly, their AABBs).  Starting with the root cell, each child cell is tested for intersection, constituting a walk through the tree.  If a leaf cell is intersected by the ray, then the particles in the leaf are tested for intersection (by calculating their impact parameters). This ensures that only a minimum fraction of the particles in the system need testing for intersection.  This illustrates the necessity of AABBs; tree nodes may contain a particle, but not its entire smoothing volume.  Thus, calculating intersections between a ray and tree nodes may miss contributions to the density field from  smoothing volumes that cross node intersections. 
  Tests for intersections between rays and AABBs are carried out using the ray slopes algorithm \citep{rayslope}, which has been shown to be faster than other commonly used methods, such as using Pl\"{u}cker coordinates \citep{plucker}.

\subsubsection{Determining the Scattering Location}

\noindent An important facet of an MCRT code is the determination of the scattering location of the photon.  In general, the scattering location will occur inside a smoothing volume, and possibly at a location where the density depends on the contributions from several particles.  Therefore, when attempting to determine the scattering location, it is important to define four classes of particle:

\begin{enumerate}
\item Particles that do not intersect the ray (\emph{unlisted})
\item Particles that intersect, but do not contain the location of emission (\emph{distant-listed})
\item Particles that intersect, and contain the location of emission in front of them (\emph{front-listed})
\item Particles that intersect, and contain the location of emssion behind them (\emph{back-listed})
\end{enumerate}

\begin{figure*}
\begin{center}$
\begin{array}{cc}
\includegraphics[scale = 0.2]{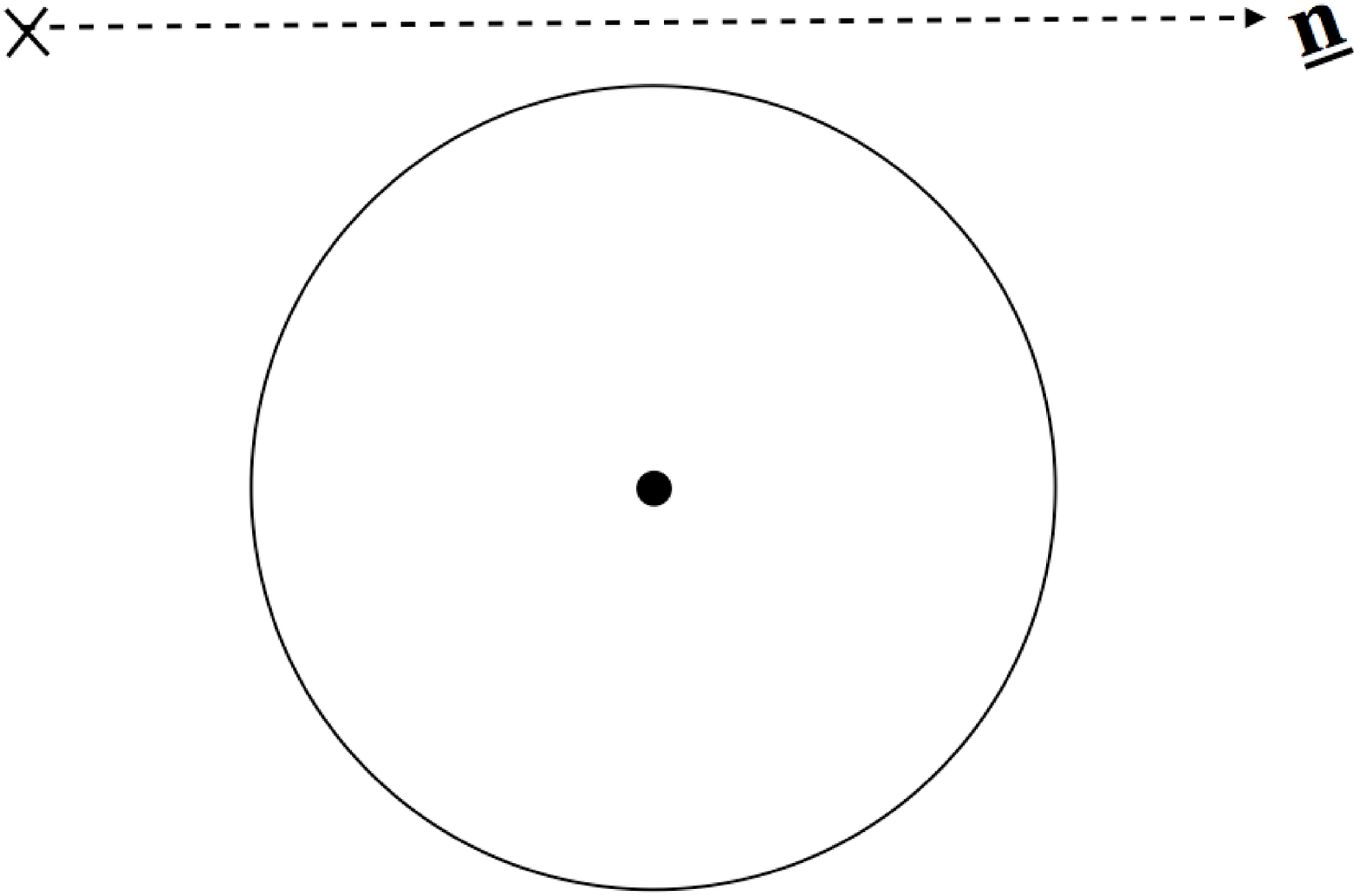} &
\includegraphics[scale = 0.2]{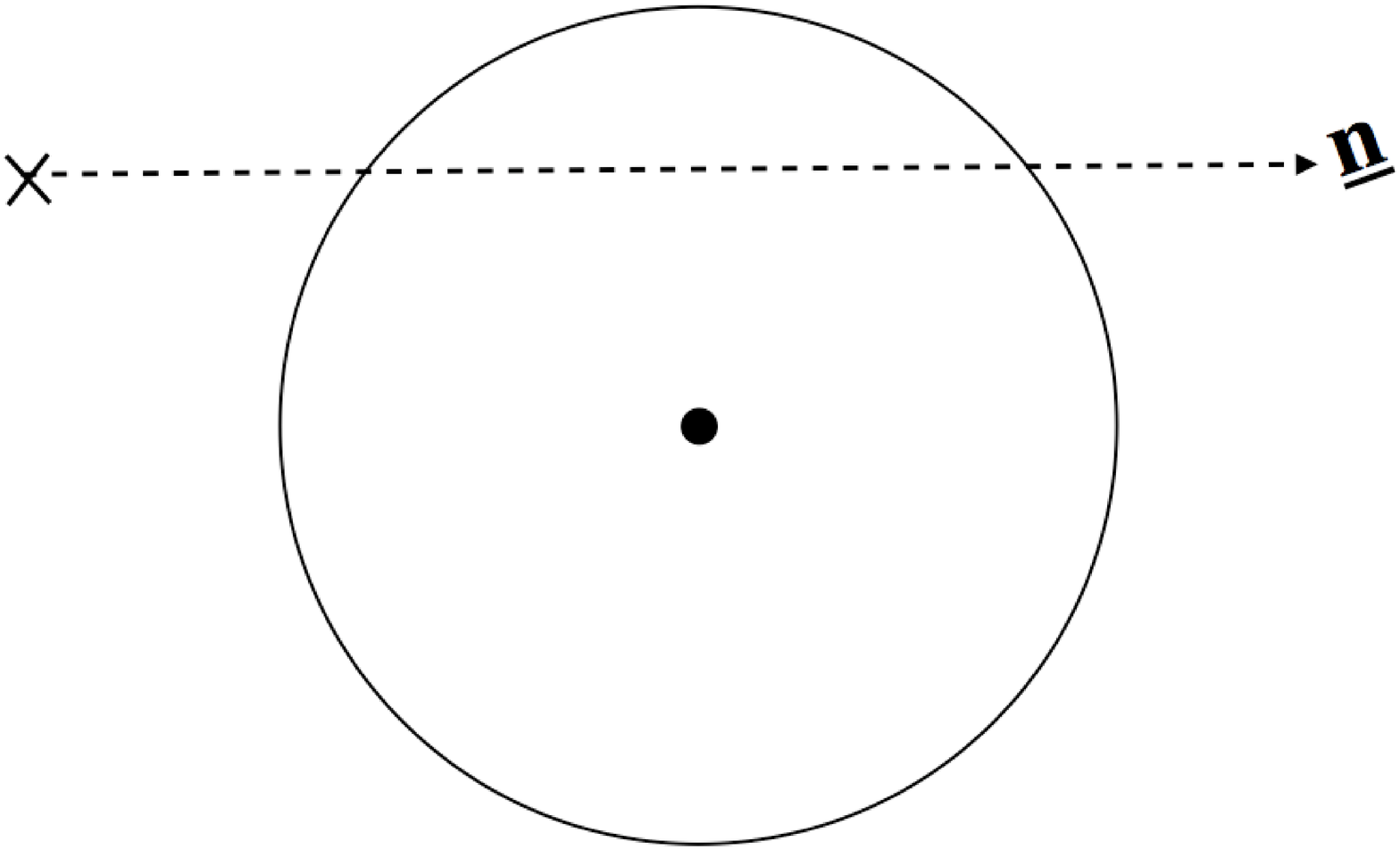} \\
\includegraphics[scale = 0.2]{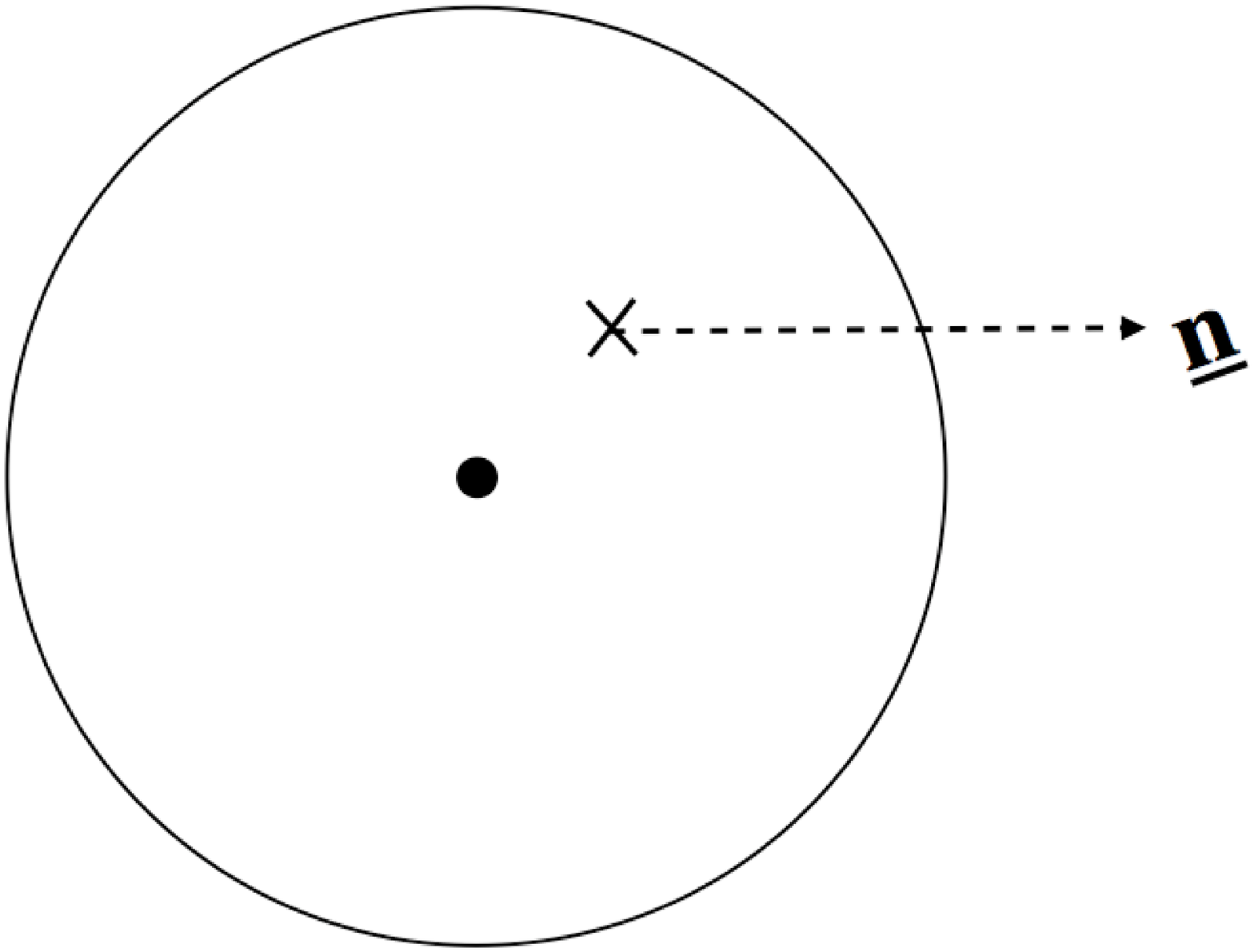} &
\includegraphics[scale = 0.2]{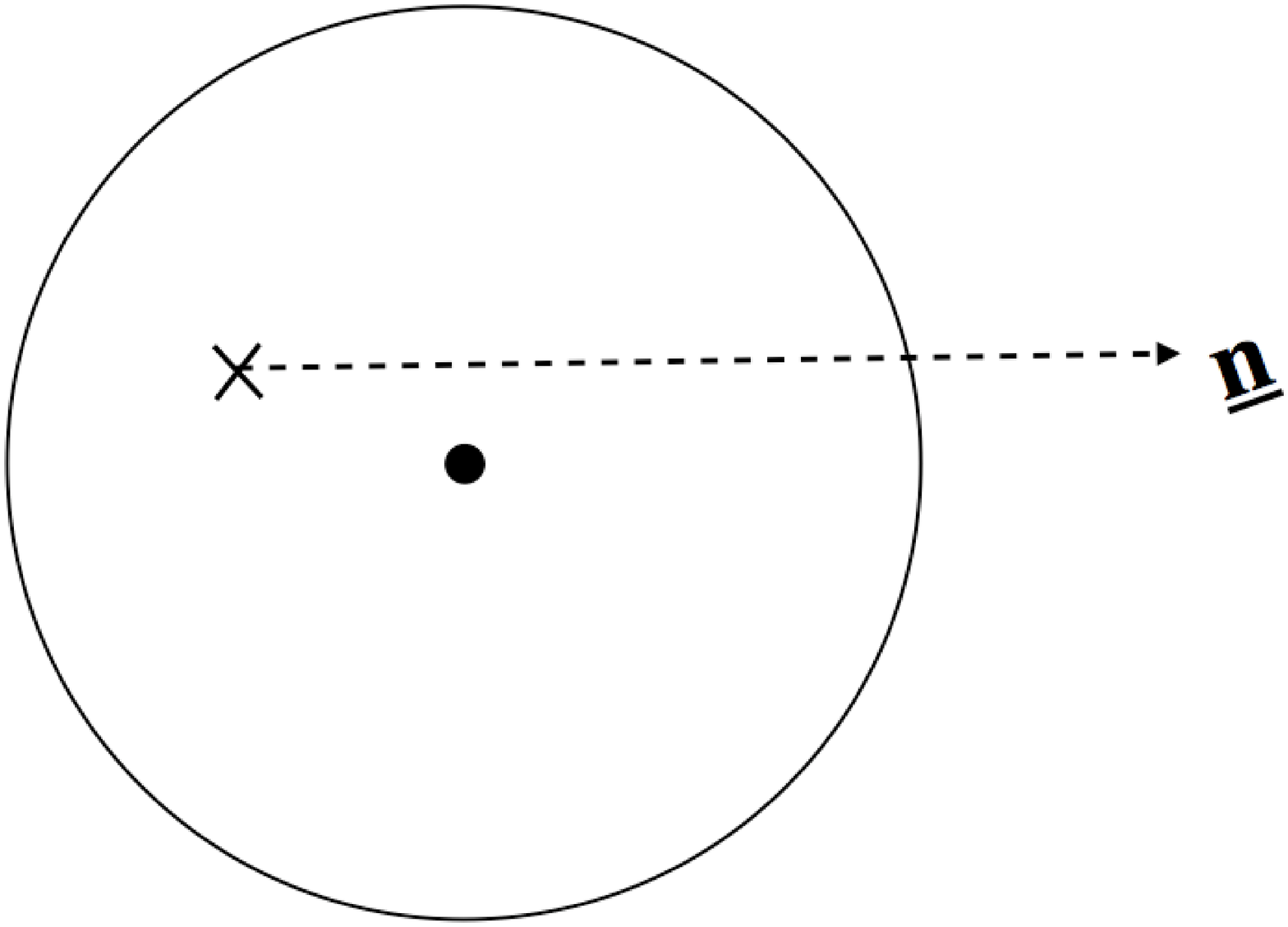} 
\end{array}$
\caption{The four classes of particle in the system: unlisted (top left), distant-listed (top right), front-listed (bottom left) and back-listed (bottom right).  The "x" denotes the emission location of the photon.\label{fig:classes}}
\end{center}
\end{figure*}	 	

\noindent The classes are illustrated in Figure \ref{fig:classes}.  Particles of class (i) obviously do not affect the calculation - particles of class (ii) are accounted for simply.  Particles of classes (iii) and (iv) will have differing effects on the optical depth calculation, and will require separate treatments.

The scattering location is determined by iteration: firstly, the optical depth is calculated particle by particle using the raylist until the optical depth exceeds the randomly selected optical depth \(\tau_{scatter}\) at particle \(k\).  Then, the optical depth is calculated from the beginning of the sphere for particle \((k-1)\) (ensuring that all potential contributors before and after this location are accounted for), iterating over distance until the answer converges on \(\tau_{scatter}\).  As the optical depth always increases with distance, convergence can be achieved with simple algorithms and relatively little computation.  This code uses a recursive bisector algorithm to perform the iteration.  Starting from the path length between the beginning of sphere $(k-1)$ to the end of sphere $k$, this value is halved recursively until the correct optical depth is obtained (to within some tolerance) or until the path length reaches a minimum value (defined as a fraction of the smallest smoothing length in the simulation).

\section{Tests and Applications}\label{sec:applications}

\subsection{Comparison with Analytic Results}

\begin{figure}
\begin{center}
\includegraphics[scale = 0.25]{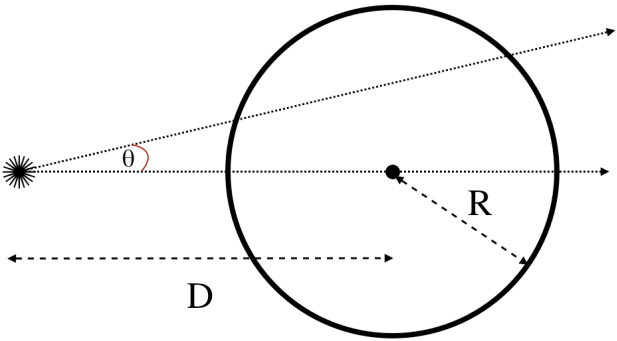}
\caption{Schematic of the raytracing experiment.  The optical depth of the ray can be calculated analytically to compare with the code's output. \label{fig:raysphere}}
\end{center}
\end{figure}

\noindent To confirm the raytracing component of the code was working correctly, a simple test case was devised (Figure \ref{fig:raysphere}).  Consider a uniform density sphere, with radius $R$, density $\rho_0$ and total opacity $\chi$.  A point source is located a distance $D$ from the centre of the sphere.  It emits rays at an angle $\theta $ from the vector connecting the source and the sphere's centre.  The optical depth $\tau(\theta)$ therefore has an analytic solution:

\begin{equation} \tau(\theta) = 2 \rho_0 \chi \sqrt{R^2 - D^2 \sin^2(\theta)} \end{equation}

\begin{figure}
\begin{center}
\includegraphics[scale = 0.5]{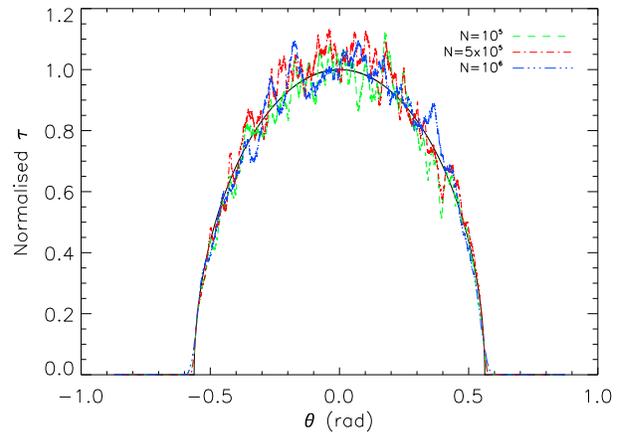}
\caption{Results of the raytracing experiment.  The black solid line indicates the analytical result; the coloured dashed lines show simulation results for increasing particle number. \label{fig:testsphere}}
\end{center}
\end{figure}	 

\noindent Three SPH snapshots were generated, each containing a uniform density sphere (mass $1\,M_{\odot}$, $R=2133$ au).  Three were generated to check convergence: snapshot 1 used $10^5$ SPH particles; snapshot 2 used $5\times 10^{5}$; snapshot 3 used $10^{6}$.  A point source was placed at a distance $D=4000$ au, and the optical depth along the ray (assuming an opacity of unity) was calculated as a function of $\theta$.  The numerical results are compared with the analytical result (scaled such that the maximum optical depth is 1) in Figure \ref{fig:testsphere}.  The column density is subject to the underlying random noise (at a level of around 5\%) associated with generating an SPH snapshot (which has not undergone any settling).  However, the results vary little with increasing particle number, showing that the column density has converged even for the relatively low particle number of $10^5$.

\subsection{A Low Mass Companion for HL Tau?}

\begin{figure*}
\begin{center}$
\begin{array}{cc}
\includegraphics[scale = 0.4]{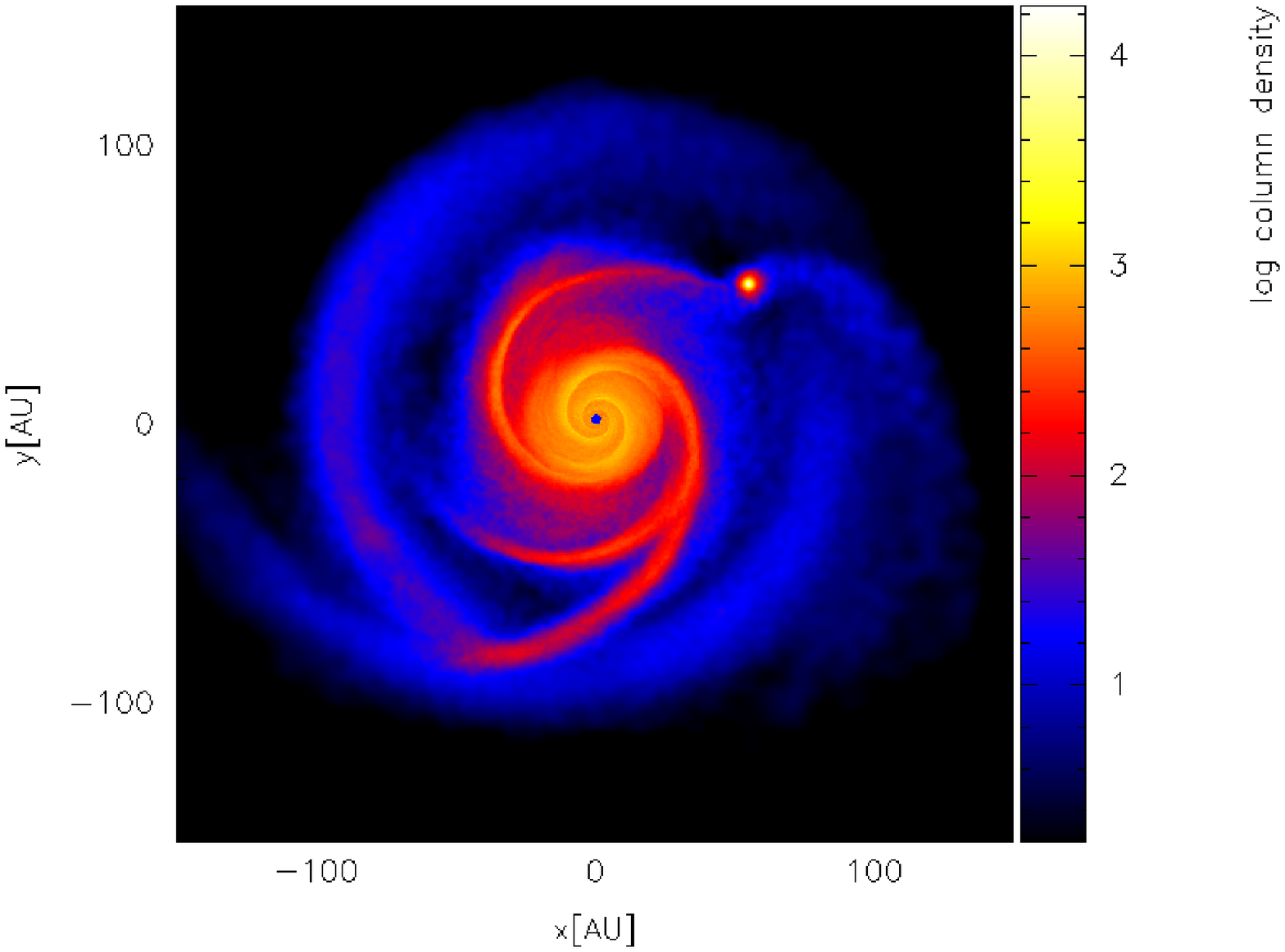} &
\includegraphics[scale = 0.4]{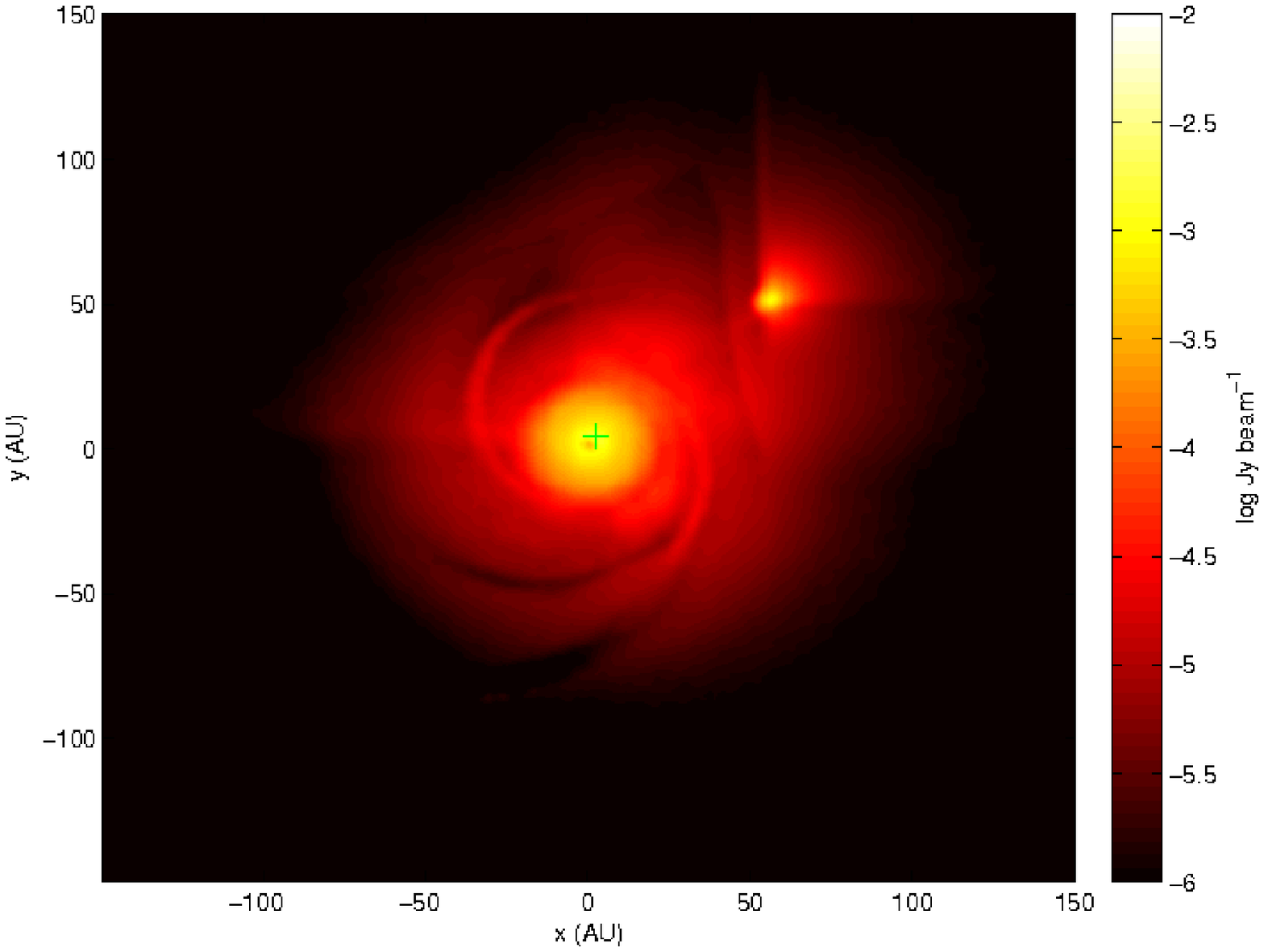} \\
\end{array}$
\caption{Left: Surface density plot of the HL Tau simulation to be imaged. Right: image of the HL Tau simulation, integrated over the wavelength range  \([0.01,0.1]\) cm.  The green cross indicates the pixel with maximum intensity. \label{fig:HL_Tau}}
\end{center}
\end{figure*}

\noindent \citet{Greaves_Tau} imaged HL Tau using the Very Large Array (VLA) with a resolution of 0.08'' at a wavelength of 1.3cm (corresponding to a spatial resolution of 10 au at HL Tau's distance of \(\sim\) 140 pc).  They discovered excess emission at \(\sim 65\) au, which they identified as a candidate protoplanet in the earliest stages of formation, a possible example of protoplanetary disc fragmentation forming a bound object \citep{Boss_science}.  To lend weight to their hypothesis, they conducted SPH simulations (containing 250,000 particles) of an unstable star-disc system with similar parameters to HL Tau, in which a clump forms at \(\sim 75\) au, with a similar mass as that deduced for the candidate (see Figure \ref{fig:HL_Tau}).  The simulation uses an SPH code based on the work of \citet{Bate_code}.  It employs individual particle timesteps, and models hydrodynamics and gravity, with radiative cooling as prescribed by \citet{Stam_2007}. What can a telescope like ALMA be expected to see in HL Tau?  Will the candidate protoplanet be resolvable at millimetre wavelengths?

The SPH simulation from \citet{Greaves_Tau} was therefore imaged, using the same dust scattering parameters as used in \citet{www03}\footnote{Note that the examples shown here approximate $\chi_{\nu} = \kappa_{\nu}$}.  The wavelength and resolution parameters were set to be representative of ALMA \footnote{http://www.eso.org/sci/facilities/alma/}: the wavelength range is \([0.01,0.1]\) cm, with an effective resolution of 0.01''.  The star is  $0.5\, M_{\odot}$, and the disc is  $0.3\, M_{\odot}$, with an initial surface density profile of $\Sigma \propto r^{-1}$  A dust to gas ratio of 0.005 (corresponding to 50\% solar metallicity) was assumed throughout. Dust sublimation is prescribed by enforcing any SPH particle with temperature greater than 1600 K to have zero dust mass.  Temperatures were calculated using the equation of state outlined in \citet{intro_hybrid} (more information can be found in \citealt{Stam_2007,Boley_hydrogen}).  The star's emission was modelled by a blackbody source with radius $2R_{\odot}$ and temperature 2000K.

Figure \ref{fig:HL_Tau} shows that the clump, which reaches temperatures of around 1500 K at its centre is detectable by its emission, although somewhat fainter than the main star-disc system.  In fact, it should be expected that the clump is fainter still, if dust sublimation is more appropriately modelled as opposed to a one temperature cut-off.  The m=2 spiral mode attached to the clump is also apparent; however with a flux of around 0.001 mJy/beam, it is unlikely that ALMA will be able to image it, highlighting the need for resolution of order $\sim$ 1 au at 100 pc \emph{and} sufficient sensitivity in the detection of disc spiral arms.  The peak emission is around 0.5 Jy, which is within a factor of two of the observed SCUBA measurements of HL Tau \footnote{http://www.jach.hawaii.edu/JCMT/continuum/calibration/sens/calibrators.html}.  The remaining discrepancy is most likely due to uncertainty in the selected disc and star parameters (as well as the dust data).  To summarise, the clump is indeed detectable with ALMA, but a telescope with better sensitivity is required to detect the spiral arms correctly. 
 
\subsection{The Observational Features of Stellar Encounters}  

\noindent It is expected that in an average stellar cluster, at least one star-disc system will undergo a close encounter with another star in its lifetime \citep{Clarke_Pringle_binary,enc_outburst}.  The effect of an encounter perturbs the disc, modifying its surface density profile and triggering enhanced accretion.  The effect of the enhanced accretion produces a stellar outburst.  This feature was  linked to the FU Orionis phenomenon \citep{Bonnell_Bastien_92,Pfalzner_08}, but it has been shown that such encounters are too infrequent to reproduce the correct observational signatures of FU Ori outbursts \citep{enc_outburst}.

\citet{encounters} carried out a series of SPH simulations to study the effects of a close encounter on a protostellar disc's dynamics and structure.  As a second application of the imaging techniques discussed here, the reference simulation described in \citet{encounters} was imaged at two points in its evolution: the initial pre-encounter phase, where the disc remains in a marginally stable, self-gravitating state \citep{Lodato_and_Rice_04}; and during periastron, where the disc is heated strongly by the secondary's motion through it.  

The primary is $0.5\, M_{\odot}$ (modelled by a blackbody source with radius $2R_{\odot}$ and temperature 2000K), and the secondary is $0.1 \, M_{\odot}$  (modelled by a blackbody source with radius $0.2R_{\odot}$ and temperature 1000K).  The disc is $0.1\, M_{\odot}$, with an initial surface density profile of $\Sigma \propto r^{-1}$ (and dust to gas ratio of 0.01, i.e metallicity equal to solar). The imaging parameters are the same as those of the HL Tau example above (\([0.01,0.1]\) cm, with an effective resolution of 0.01'').

\begin{figure*}
\begin{center}$
\begin{array}{cc}
\includegraphics[scale = 0.4]{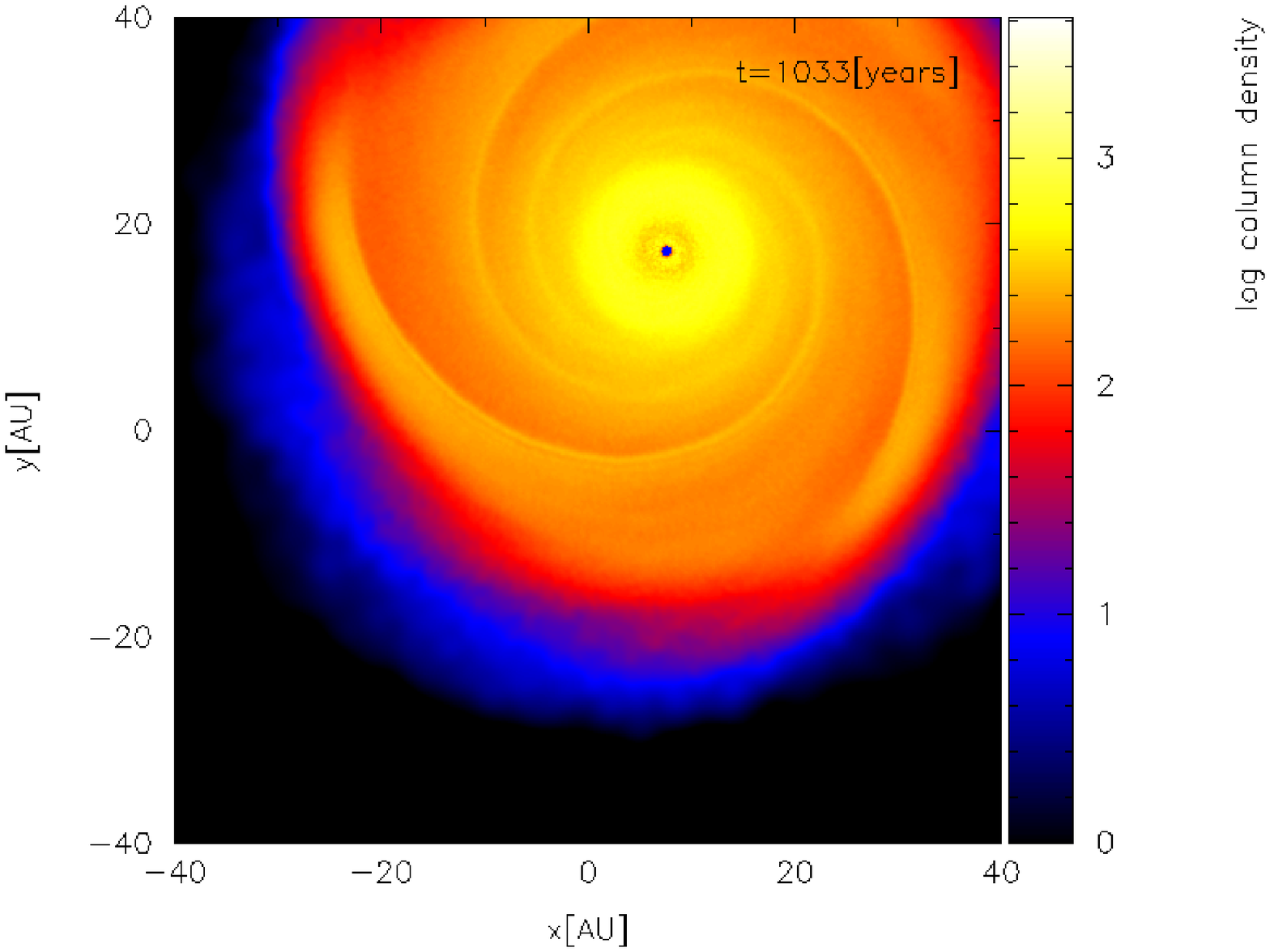} &
\includegraphics[scale = 0.4]{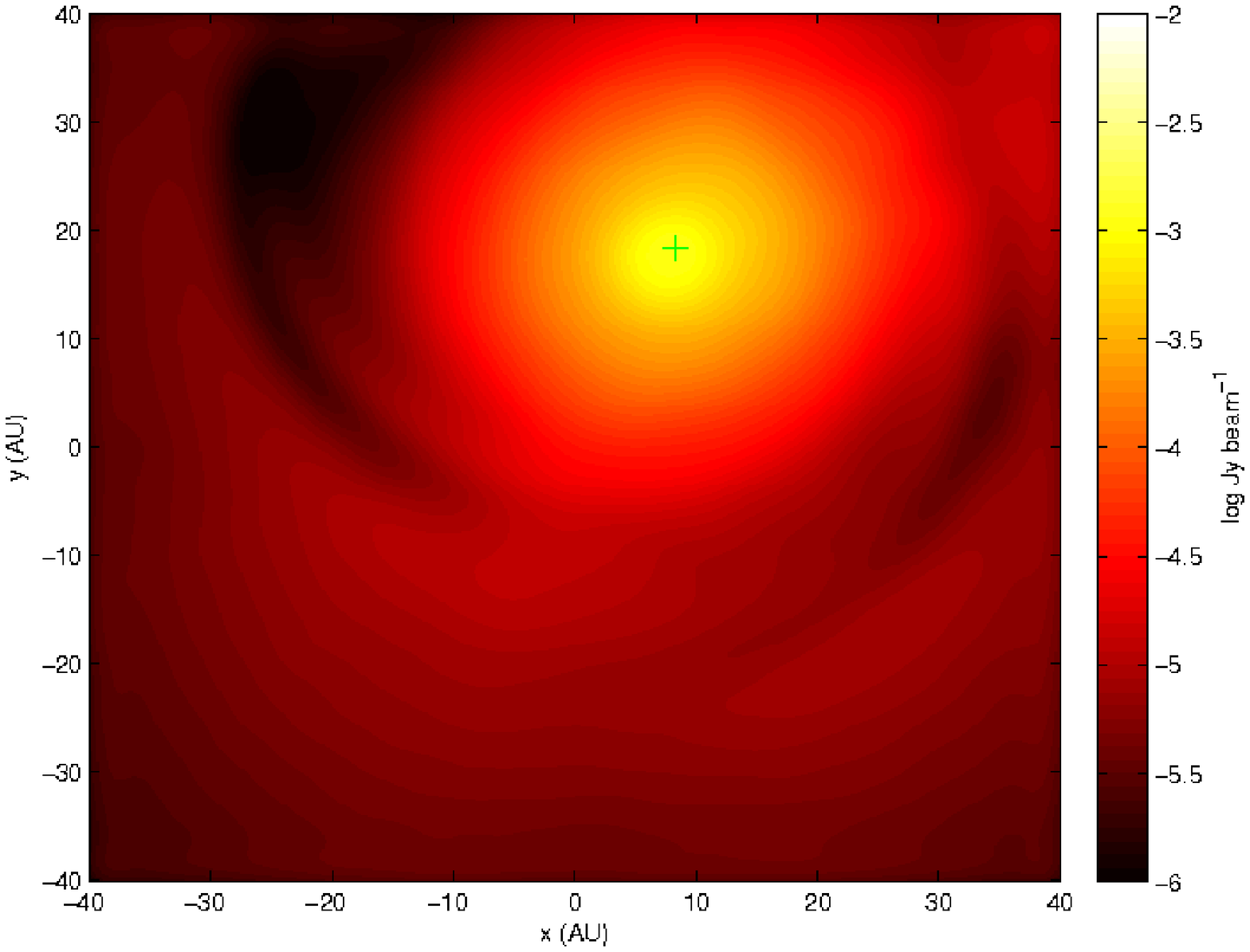} \\
\end{array}$
\caption{Left: Surface density plot of the disc pre-outburst.  The secondary is out of frame. Right: image of the simulation, integrated over the wavelength range  \([0.01,0.1]\) cm.  The green cross indicates the pixel with maximum intensity \label{fig:quiescent}}
\end{center}
\end{figure*}

\begin{figure*}
\begin{center}$
\begin{array}{cc}
\includegraphics[scale = 0.4]{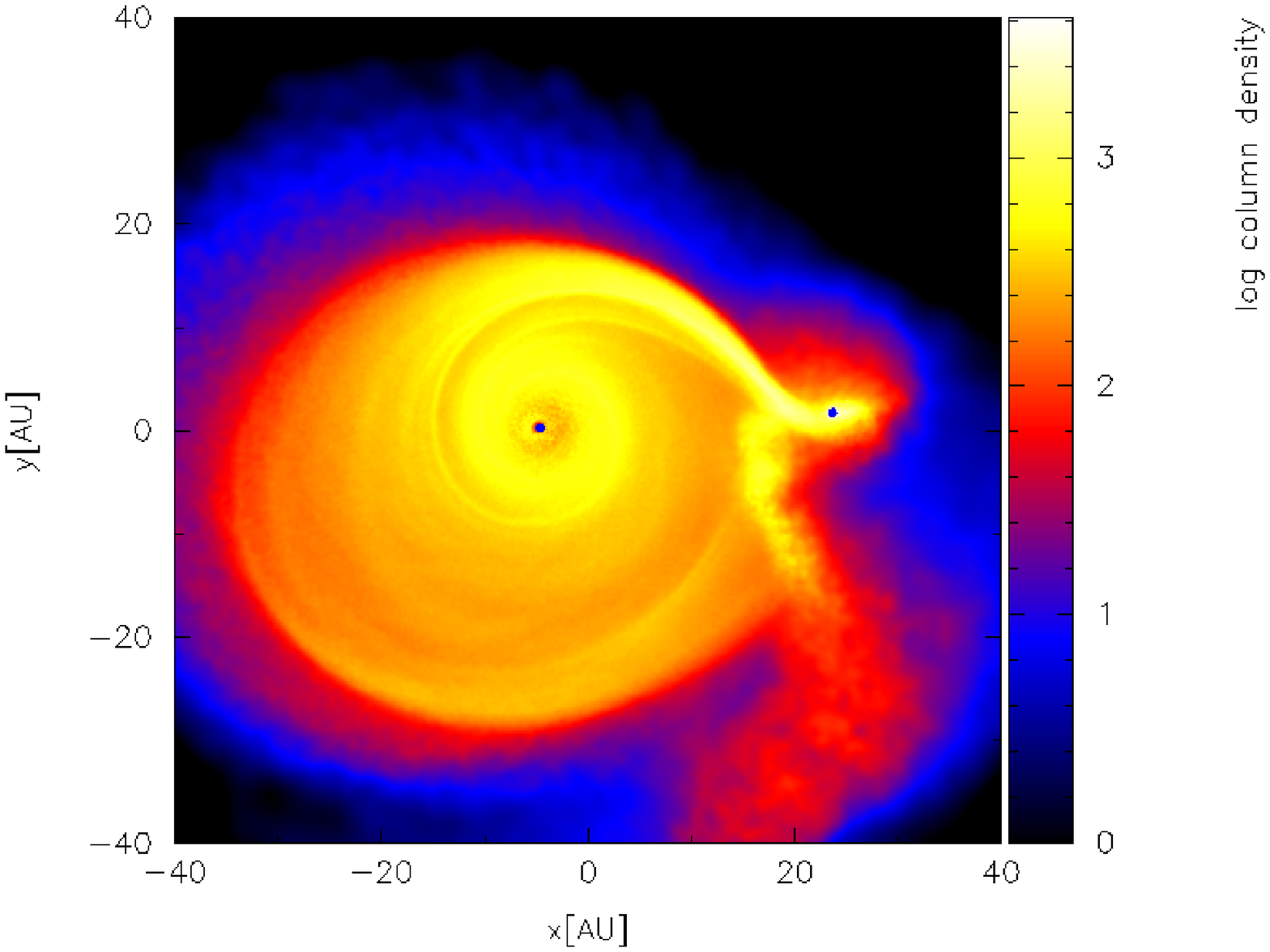} &
\includegraphics[scale = 0.4]{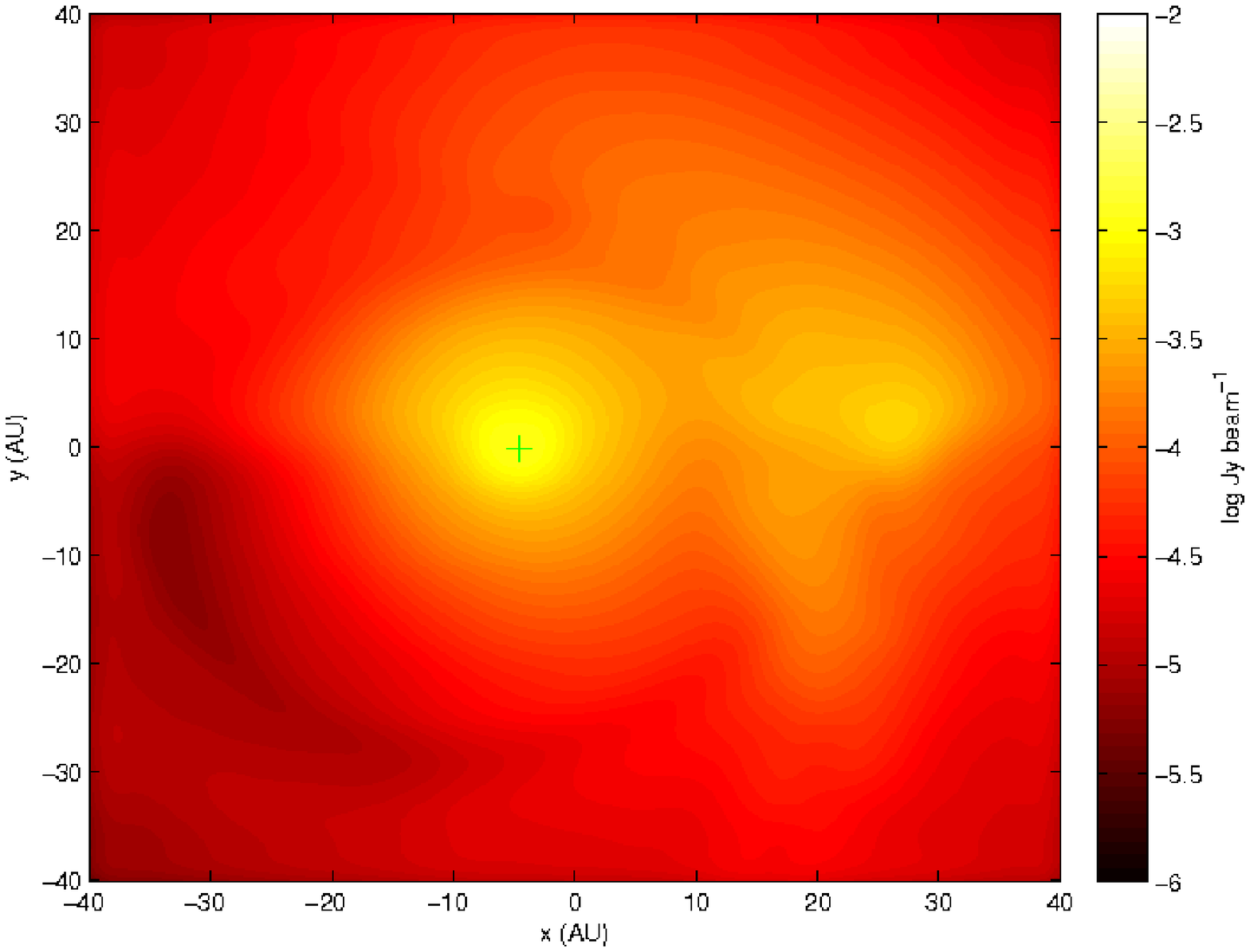} \\
\end{array}$
\caption{Left: Surface density plot of the disc at periastron. Right: image of the simulation, integrated over the wavelength range  \([0.01,0.1]\) cm.  The green cross indicates the pixel with maximum intensity \label{fig:periastron}}
\end{center}
\end{figure*}

Disc asymmetry plays an important role in the imaging of this system (Figures \ref{fig:quiescent} and \ref{fig:periastron}).  In Figure \ref{fig:quiescent}, the disc displays a non-zero ellipticity due to its spiral structure, with shadowing indicating the increased surface density of the disc corresponding to the spirals themselves.  Again, however, these shadows are only apparent if the sensitivity of the telescope is 1 $\mu$Jy/beam, well beyond the ALMA's current continuum sensitivity estimates.  This ellipticity is enhanced during the encounter (Figure \ref{fig:periastron}), with the semi major axis of the ellipse aligned with the orbital vector of the primary and secondary.  The erasure of the strong spiral structure during the encounter results in a detectable tidal arm with flux around 1 mJy/beam.  The stellar emission is dwarfed by the disc at these wavelengths.  The inner regions of the disc are not hot enough for dust sublimation to open a resolvable gap (the intrinsic inner gap is very much below the resolution limit).  Again, the resolution of spiral structure in a self-gravitating disc will not be possible with ALMA unless higher sensitivities are achieved, but it \emph{will} be possible to detect enhanced emission from a tidal spiral wave generated as the result of a stellar encounter. 

\section{Discussion}\label{sec:discussion}

\subsection{Runtime Scaling}

\noindent As the SPH systems being imaged by this code will be in general disordered and impossible to render analytically, a true runtime scaling is not feasible.  However, an example scaling assuming simple geometry can be calculated.

In general, the runtime \(T\) goes as

\begin{equation} T \sim N_{\gamma} N_{steps} \end{equation}

Where \(N_{\gamma}\) is the number of photons emitted by the code, and \(N_{steps}\) represents the computational expense required to track one photon from emission to capture.  This can hence be written

\begin{equation} T \sim N_{\gamma} <N_{ray}> N_{scatt} \end{equation}

Where \(<N_{ray}>\) is the mean number of particles intersected by any ray, and \(N_{scatt}\) indicates how many times a photon will scatter before it exits.  Generally, \(N_{scatt} \sim <\tau>^2\), and

\begin{equation} <N_{ray}> \sim N_{p}\, P_{intersect} \end{equation}

Where \(N_{p}\) is the number of SPH particles, and \(P_{intersect}\) is the probability that any one particle is intersected by the ray.  This can be estimated for simple geometries: assuming the SPH particle distribution is a sphere, then

\begin{equation} P_{intersect} \sim 1 - e^{-Rn\sigma_s} \end{equation}

Where R is the sphere's radius,  \(n = N_{p}/V \sim N_{p}/R^3\) is the number density of the sphere, and \(\sigma_s \sim 4 \pi <h>^2\) is the average cross-section of the SPH particles.  In a homogeneous sphere, 

\begin{equation} <h> \sim \left(\frac{1}{\rho}\right)^{\frac{1}{3}} \sim \frac{R}{N_{p}^{\frac{1}{3}}} \end{equation}

Combining these results gives

\begin{equation} T \sim N_{\gamma} <\tau>^2 N_{p}\, \left(1 - e^{-N_{p}^{\frac{1}{3}}}\right) \end{equation}

\noindent This runtime scaling shares common features with grid-based MCRT codes, in particular the dependence on optical depth and number of particles/cells. For illustration, the simulations used in this work typically took approximately 100 CPU hours to create an image, with $N_{p}=500,000$, $N_{gamma} = 10^8$ in systems with significant optically thick components.

\subsection{Resolution \label{sec:res}}

\noindent SPH simulations are typically subject to several resolution conditions, principal among them the Jeans Resolution criterion of \citet{Burkert_Jeans} to correctly resolve fragmentation.  Imaging these simulations is no different - it must be demonstrable that the local density field is sufficiently resolved to allow a satisfactory calculation of the optical depth (more specifically the optical depth to the photosphere, as this is where most of the received emission will originate).  In terms of the smoothing length, $h$, and the mean free path $\lambda_{mfp}$, this condition is

\begin{equation} h < \lambda_{mfp} = \frac{1}{\rho \kappa}. \end{equation}

\noindent As the value of $h$ is inversely proportional to the local number density of SPH particles, the resolution of an MCRT image of an SPH density field is closely linked to the total number of particles in the simulation, an intuitive result.  This condition is satisfied in the outer disc easily, as the mean free path is of order the system scale height, (which is resolved by several smoothing lengths).  However, within the inner 5-10 au of the disc, two separate issues arise:

\begin{enumerate}
\item The smoothing length exceeds the local mean free path.  This is partially compensated by SPH's smoothing prescription - the density field is defined continously across all space, and therefore affords a kind of sub-$h$ resolution.  However, this smoothing erases information about fluctuations in the density field below this length scale, and this will lead to an overestimation of the escaping flux.
\item The disc is not well resolved several scale heights above the midplane, where the photosphere is located.  This under-resolving will result in a decrease in scattering events at the photosphere, affecting the scattered light component of the flux.  
\end{enumerate}

These two issues show that inside $\sim 10$ au, the flux escaping from the disc will be subject to resolution-based error.   Users concerned with the inner regions of these discs could adopt a particle splitting prescription to boost the resolution (e.g. \citealt{SPH_split}).   However, as this work is concerned mainly with millimetre emission from the cooler, well-resolved outer disc, this emission will be adequately resolved, and the issues described above will not play a significant role. 

\subsection{Radiative Equilibrium: A Proof of Concept}

\noindent The work described here has directly utilised the temperature structures produced in the SPH simulations, but such outputs are not a requirement for imaging them using MCRT.  The temperature for each fluid element can be calculated using MCRT with so-called radiative equilibrium methods (e.g. \citealt{Bjorkman_Wood_MCRE}).  In essence, these methods involve absorbed photons incrementing the local temperature field at the absorption point \(T(\mathbf{r}_j)\) by an amount \(\Delta T(\mathbf{r}_j)\), and then being re-emitted.  As many photons are passed through the system and are absorbed and re-emitted at various locations, the temperature structure will relax to the correct value.  In grid-based methods, the procedure of increasing the local temperature is simple - the increment \(\Delta T\) is added to the grid cell the photon is absorbed in.  In SPH fields, the situation is somewhat more complex. A scalar field \(X\) at a location \(\mathbf{r}_j\) is defined as:

\begin{equation} X(\textbf{r}_j) = \sum_{i} \frac{X_im_i}{\rho_i} \textbf{W}(\textbf{r}_i - \textbf{r}_j,h_i) \end{equation}

\noindent (Where we have used the ``scatter" interpretation).  We can substitute for the temperature increase \(\Delta T\):

\begin{equation} \Delta T(\textbf{r}_j) = \sum_{i} \frac{\Delta T_{ij}m_i}{\rho_i} \textbf{W}(\textbf{r}_i - \textbf{r}_j,h_i) \label{eq:sphre}\end{equation}

\noindent The left hand side of equation (\ref{eq:sphre}) can be calculated using the traditional radiative equilibrium methods, and is known.  The desired quantities are the \(\Delta T_{ij}\), that is the individual increases in temperature each SPH particle must be assigned.  Equation (\ref{eq:sphre}) therefore acts as a constraint on the \(\Delta T_{ij}\), but is insufficient to close the system.  A method of closure is presented here as proof that the radiative equilibrium framework can indeed be incorporated - its application is left for future work.  Closure can be achieved by specifying the following \emph{ansatz}:

\begin{equation} \Delta T_{ij} = A_{ij}\Delta T g(r_{ij}) \end{equation}

\noindent Where \(A_{ij}\) is some real number, \(r_{ij}\) is the separation of particle \(i\) from the location \(j\) and \(g(r_{ij})\) is some function that satisfies:

\begin{equation} g(0) = 1, \end{equation}
\begin{equation} g(r_{ij} > 2h_{i}) = 0 \end{equation}

\noindent i.e. particles with smoothing volumes that do not intersect the location \(j\) do not contribute.  This will satisfy equation (\ref{eq:sphre}) provided

\begin{equation} \sum_{i} \frac{m_i}{\rho_i}A_{ij} g(r_{ij}) \textbf{W}({r}_{ij},h_i) = 1 \end{equation}

\noindent This has recast the problem from solving for \(\Delta T_{ij}\) to the prefactors \(A_{ij}\) and function $g(r_{ij})$.  This may appear to be a sideways step: however, giving a specific example is illuminating.  To obtain a suitable function \(g(r_{ij})\), consider the decay of flux \(F_0\) along a path of optical depth \(\tau\):

\begin{equation} F = F_0 e^{-\tau} \end{equation}

\noindent Using the Stefan-Boltzmann law, \(F = \sigma T^4\) yields

\begin{equation} \sigma (\Delta T_{ij})^4 = \sigma (\Delta T)^4 e^{-\tau_{ij}} \end{equation}

\noindent Inspecting the above equation suggests the following form for $g$:

\begin{equation} g(r_{ij}) = e^{-\tau_{ij}/4} \end{equation}

\noindent Assuming that \(A_{ij} = A\) for all (\(i,j\)), the normalisation condition becomes

\begin{equation} A = \frac{1}{\sum_{i} \frac{m_i}{\rho_i}e^{-\tau_{ij}/4} \textbf{W}({r}_{ij},h_i)} \end{equation}

\noindent The system has now been closed with an \emph{ansatz} that is physically motivated, and reproduces expected behaviour (particles with a reduced optical depth to the location will receive a greater temperature boost than others in more optically thick regions).  

Other radiative equilibrium methods can also be recast in SPH form, such as that of \citet{Lucy_RE}, which computes grid cell temperatures by counting the path lengths of any photons that pass through it during one iteration, retrieving the temperature structure rapidly within a few iterations.  By replacing grid cells with particle smoothing volumes, it can be seen that this method is indeed amenable to SPH fields also.

\section{Conclusions}\label{sec:conclusions}

\noindent This paper has outlined a means for applying Monte Carlo Radiative Transfer (MCRT) techniques directly to Smoothed Particle Hydrodynamics (SPH) density fields of star and planet formation.  In doing so, it gives a means for synthetic telescope images to be made, allowing the flexibility of SPH simulations to be retained in calculating optical depths, scattering and polarisation.  This work has shown two applications of the method to SPH simulations of protostellar discs: non-trivial features in the imaging of these environments are well-traced, and the resulting images give new opportunities for theoretical input to observed discs.  It can be shown what resolutions and sensitivities are required to observe these features.  ALMA appears to be able to resolve and image perturbed structures like tidal arms in discs undergoing close encounters.  But, while it has sufficient resolution, it cannot image unperturbed spiral structure (or the shadows it casts) without an improvement in sensitivity by at least a factor of 100.  Its inherently graphical nature allows the code to be greatly optimised, whether by standard parallelisation methods or by the use of Graphical Processing Units (GPUs) to handle the ray intersections.

While used for imaging in this work, the method is not restricted to this alone.  As the algorithm only modifies how the optical depth is calculated (in order to work in SPH fields), it can be applied to any circumstances traditional gridded MCRT has been applied to.  This includes radiative equilibrium simulations where the temperature structure is calculated from stellar emission, or moving beyond continuum emission to perform line radiative transfer calculations.  In effect this places SPH on a par with grids or meshes for MCRT techniques.

\section*{Acknowledgements}

The authors would like to thank Matthew Bate and Barbara Whitney for useful comments which greatly strengthened this work. Plots of the SPH simulations were made using \begin{small}{SPLASH}\end{small} \citep{SPLASH}.  All simulations presented in this work were carried out using high performance computing funded by the Scottish Universities Physics Alliance (SUPA).

\bibliographystyle{mn2e} 
\bibliography{mcsph}

\begin{thebibliography}{}

\bibitem[\protect\citeauthoryear{{Altay}, {Croft} \& {Pelupessy}}{{Altay}
  et~al.}{2008}]{SPHRAY}
{Altay} G.,  {Croft} R.~A.~C.,    {Pelupessy} I.,  2008, \mnras, 386, 1931

\bibitem[\protect\citeauthoryear{{Bastien}, {Cha} \& {Viau}}{{Bastien}
  et~al.}{2004}]{Bastien_diffusion}
{Bastien} P.,  {Cha} S.,    {Viau} S.,  2004, in {G.~Garcia-Segura,
  G.~Tenorio-Tagle, J.~Franco, \& H.~W.~Yorke } ed., Revista Mexicana de
  Astronomia y Astrofisica Conference Series Vol.~22 of Revista Mexicana de
  Astronomia y Astrofisica Conference Series, {SPH with radiative transfer:
  method and applications}.
pp 144--147

\bibitem[\protect\citeauthoryear{{Bate}, {Bonnell} \& {Price}}{{Bate}
  et~al.}{1995}]{Bate_code}
{Bate} M.~R.,  {Bonnell} I.~A.,    {Price} N.~M.,  1995, \mnras, 277, 362

\bibitem[\protect\citeauthoryear{{Bate} \& {Burkert}}{{Bate} \&
  {Burkert}}{1997}]{Burkert_Jeans}
{Bate} M.~R.,  {Burkert} A.,  1997, \mnras, 288, 1060

\bibitem[\protect\citeauthoryear{{Bisbas}, {W{\"u}nsch}, {Whitworth} \&
  {Hubber}}{{Bisbas} et~al.}{2009}]{Bisbas_et_al_09}
{Bisbas} T.~G.,  {W{\"u}nsch} R.,  {Whitworth} A.~P.,    {Hubber} D.~A.,  2009,
  \aap, 497, 649

\bibitem[\protect\citeauthoryear{{Bjorkman} \& {Wood}}{{Bjorkman} \&
  {Wood}}{2001}]{Bjorkman_Wood_MCRE}
{Bjorkman} J.~E.,  {Wood} K.,  2001, \apj, 554, 615

\bibitem[\protect\citeauthoryear{{Boley}, {Hartquist}, {Durisen} \&
  {Michael}}{{Boley} et~al.}{2007}]{Boley_hydrogen}
{Boley} A.~C.,  {Hartquist} T.~W.,  {Durisen} R.~H.,    {Michael} S.,  2007,
  \apjl, 656, L89

\bibitem[\protect\citeauthoryear{{Bonnell} \& {Bastien}}{{Bonnell} \&
  {Bastien}}{1992}]{Bonnell_Bastien_92}
{Bonnell} I.,  {Bastien} P.,  1992, \apjl, 401, L31

\bibitem[\protect\citeauthoryear{{Boss}}{{Boss}}{1997}]{Boss_science}
{Boss} A.~P.,  1997, Science, 276, 1836

\bibitem[\protect\citeauthoryear{{Clarke} \& {Pringle}}{{Clarke} \&
  {Pringle}}{1991}]{Clarke_Pringle_binary}
{Clarke} C.~J.,  {Pringle} J.~E.,  1991, \mnras, 249, 584

\bibitem[\protect\citeauthoryear{Eisemann, Grosch, Müller \& Magnor}{Eisemann
  et~al.}{2007}]{rayslope}
Eisemann M.,  Grosch T.,  Müller S.,    Magnor M.,  2007, J. Graphics Tools,
  12, 35

\bibitem[\protect\citeauthoryear{{Forgan} \& {Rice}}{{Forgan} \&
  {Rice}}{2009a}]{encounters}
{Forgan} D.,  {Rice} K.,  2009a, \mnras, 400, 2022

\bibitem[\protect\citeauthoryear{{Forgan} \& {Rice}}{{Forgan} \&
  {Rice}}{2009b}]{enc_outburst}
{Forgan} D.,  {Rice} K.,  2009b, \mnras, in press

\bibitem[\protect\citeauthoryear{{Forgan}, {Rice}, {Stamatellos} \&
  {Whitworth}}{{Forgan} et~al.}{2009}]{intro_hybrid}
{Forgan} D.,  {Rice} K.,  {Stamatellos} D.,    {Whitworth} A.,  2009, \mnras,
  394, 882

\bibitem[\protect\citeauthoryear{{Gingold} \& {Monaghan}}{{Gingold} \&
  {Monaghan}}{1977}]{Gingold_Monaghan}
{Gingold} R.~A.,  {Monaghan} J.~J.,  1977, \mnras, 181, 375

\bibitem[\protect\citeauthoryear{{Greaves}, {Richards}, {Rice} \&
  {Muxlow}}{{Greaves} et~al.}{2008}]{Greaves_Tau}
{Greaves} J.~S.,  {Richards} A.~M.~S.,  {Rice} W.~K.~M.,    {Muxlow} T.~W.~B.,
  2008, \mnras, 391, L74

\bibitem[\protect\citeauthoryear{{Hernquist} \& {Katz}}{{Hernquist} \&
  {Katz}}{1989}]{Hernquist_and_Katz_89}
{Hernquist} L.,  {Katz} N.,  1989, \apjs, 70, 419

\bibitem[\protect\citeauthoryear{{Hosking} \& {Whitworth}}{{Hosking} \&
  {Whitworth}}{2004}]{Hosking_Whitworth_04}
{Hosking} J.~G.,  {Whitworth} A.~P.,  2004, \mnras, 347, 994

\bibitem[\protect\citeauthoryear{{Kessel-Deynet} \& {Burkert}}{{Kessel-Deynet}
  \& {Burkert}}{2000}]{Kessel_Deynet_and_Burkert_00}
{Kessel-Deynet} O.,  {Burkert} A.,  2000, \mnras, 315, 713

\bibitem[\protect\citeauthoryear{{Kitsionas} \& {Whitworth}}{{Kitsionas} \&
  {Whitworth}}{2002}]{SPH_split}
{Kitsionas} S.,  {Whitworth} A.~P.,  2002, \mnras, 330, 129

\bibitem[\protect\citeauthoryear{{Kurosawa}, {Harries}, {Bate} \&
  {Symington}}{{Kurosawa} et~al.}{2004}]{Kurosawa_et_al_04}
{Kurosawa} R.,  {Harries} T.~J.,  {Bate} M.~R.,    {Symington} N.~H.,  2004,
  \mnras, 351, 1134

\bibitem[\protect\citeauthoryear{{Lodato} \& {Rice}}{{Lodato} \&
  {Rice}}{2004}]{Lodato_and_Rice_04}
{Lodato} G.,  {Rice} W.~K.~M.,  2004, \mnras, 351, 630

\bibitem[\protect\citeauthoryear{{Lucy}}{{Lucy}}{1977}]{Lucy}
{Lucy} L.~B.,  1977, \aj, 82, 1013

\bibitem[\protect\citeauthoryear{{Lucy}}{{Lucy}}{1999}]{Lucy_RE}
{Lucy} L.~B.,  1999, \aap, 344, 282

\bibitem[\protect\citeauthoryear{Mahovsky \& Wyvill}{Mahovsky \&
  Wyvill}{2004}]{plucker}
Mahovsky J.,  Wyvill B.,  2004, J. Graphics Tools, 9, 35

\bibitem[\protect\citeauthoryear{{Mayer}, {Lufkin}, {Quinn} \&
  {Wadsley}}{{Mayer} et~al.}{2007}]{Mayer_et_al_07}
{Mayer} L.,  {Lufkin} G.,  {Quinn} T.,    {Wadsley} J.,  2007, \apjl, 661, L77

\bibitem[\protect\citeauthoryear{{Monaghan}}{{Monaghan}}{1992}]{Monaghan_92}
{Monaghan} J.~J.,  1992, \araa, 30, 543

\bibitem[\protect\citeauthoryear{{Monaghan}}{{Monaghan}}{2005}]{Monaghan_05}
{Monaghan} J.~J.,  2005, \repprog, 68, 1703

\bibitem[\protect\citeauthoryear{{Oxley} \& {Woolfson}}{{Oxley} \&
  {Woolfson}}{2003}]{Oxley_Woolfson_2003}
{Oxley} S.,  {Woolfson} M.~M.,  2003, \mnras, 343, 900

\bibitem[\protect\citeauthoryear{{Pawlik} \& {Schaye}}{{Pawlik} \&
  {Schaye}}{2008}]{TRAPHIC}
{Pawlik} A.~H.,  {Schaye} J.,  2008, \mnras, 389, 651

\bibitem[\protect\citeauthoryear{{Pfalzner}}{{Pfalzner}}{2008}]{Pfalzner_08}
{Pfalzner} S.,  2008, \aap, 492, 735

\bibitem[\protect\citeauthoryear{{Pinte}, {M{\'e}nard}, {Duch{\^e}ne} \&
  {Bastien}}{{Pinte} et~al.}{2006}]{MCFOST}
{Pinte} C.,  {M{\'e}nard} F.,  {Duch{\^e}ne} G.,    {Bastien} P.,  2006, \aap,
  459, 797

\bibitem[\protect\citeauthoryear{{Price}}{{Price}}{2007}]{SPLASH}
{Price} D.~J.,  2007, Publications of the Astronomical Society of Australia,
  24, 159

\bibitem[\protect\citeauthoryear{{Price} \& {Bate}}{{Price} \&
  {Bate}}{2007}]{Price_and_Bate_07}
{Price} D.~J.,  {Bate} M.~R.,  2007, \mnras, 377, 77

\bibitem[\protect\citeauthoryear{{Price} \& {Monaghan}}{{Price} \&
  {Monaghan}}{2004}]{Price_Monaghan_SPMHD2}
{Price} D.~J.,  {Monaghan} J.~J.,  2004, \mnras, 348, 139

\bibitem[\protect\citeauthoryear{{Price} \& {Monaghan}}{{Price} \&
  {Monaghan}}{2005}]{Price_Monaghan_SPMHD3}
{Price} D.~J.,  {Monaghan} J.~J.,  2005, \mnras, 364, 384

\bibitem[\protect\citeauthoryear{{Rice}, {Armitage}, {Bate} \&
  {Bonnell}}{{Rice} et~al.}{2003}]{Ken_1}
{Rice} W.~K.~M.,  {Armitage} P.~J.,  {Bate} M.~R.,    {Bonnell} I.~A.,  2003,
  \mnras, 339, 1025

\bibitem[\protect\citeauthoryear{{Stamatellos} \& {Whitworth}}{{Stamatellos} \&
  {Whitworth}}{2005}]{Stam_MCRT}
{Stamatellos} D.,  {Whitworth} A.~P.,  2005, \aap, 439, 153

\bibitem[\protect\citeauthoryear{{Stamatellos}, {Whitworth}, {Bisbas} \&
  {Goodwin}}{{Stamatellos} et~al.}{2007}]{Stam_2007}
{Stamatellos} D.,  {Whitworth} A.~P.,  {Bisbas} T.,    {Goodwin} S.,  2007,
  \aap, 475, 37

\bibitem[\protect\citeauthoryear{{Viau}, {Bastien} \& {Cha}}{{Viau}
  et~al.}{2006}]{Viau_et_al_06}
{Viau} S.,  {Bastien} P.,    {Cha} S.,  2006, \apj, 639, 559

\bibitem[\protect\citeauthoryear{{Whitehouse} \& {Bate}}{{Whitehouse} \&
  {Bate}}{2004}]{WB_1}
{Whitehouse} S.~C.,  {Bate} M.~R.,  2004, \mnras, 353, 1078

\bibitem[\protect\citeauthoryear{{Whitehouse} \& {Bate}}{{Whitehouse} \&
  {Bate}}{2006}]{WB_2}
{Whitehouse} S.~C.,  {Bate} M.~R.,  2006, \mnras, 367, 32

\bibitem[\protect\citeauthoryear{{Whitney} \& {Hartmann}}{{Whitney} \&
  {Hartmann}}{1992}]{Whitney_and_Hartmann_92}
{Whitney} B.~A.,  {Hartmann} L.,  1992, \apj, 395, 529

\bibitem[\protect\citeauthoryear{{Wood}, {Bjorkman}, {Whitney} \&
  {Code}}{{Wood} et~al.}{1996}]{Wood_et_al_96}
{Wood} K.,  {Bjorkman} J.~E.,  {Whitney} B.~A.,    {Code} A.~D.,  1996, \apj,
  461, 828

\bibitem[\protect\citeauthoryear{{Wood}, {Wolff}, {Bjorkman} \&
  {Whitney}}{{Wood} et~al.}{2002}]{www03}
{Wood} K.,  {Wolff} M.~J.,  {Bjorkman} J.~E.,    {Whitney} B.,  2002, \apj,
  564, 887

\end{thebibliography}

\appendix

\label{lastpage}

\end{document}